\documentclass[
    aps, prx,
    floatfix,
    twoside,
    twocolumn,
    10pt,
    citeautoscript,
    superscriptaddress,
]{revtex4-2}

\usepackage[version=4]{mhchem}
\usepackage{amsmath}
\usepackage{amssymb}
\usepackage{booktabs}
\usepackage{comment}
\usepackage{glossaries}
\usepackage{graphicx}
\usepackage{tabularx}
\usepackage{siunitx}
\usepackage[dvipsnames]{xcolor}
\usepackage{float}
\usepackage[colorlinks=true]{hyperref}
\usepackage{cleveref}
\hypersetup{
    pdffitwindow=false,
    pdfstartview={FitH},
    pdfnewwindow=true,
    colorlinks=true,
    linkcolor=NavyBlue,
    citecolor=NavyBlue,
    filecolor=NavyBlue,
    urlcolor=NavyBlue,
}

\setacronymstyle{long-short}
\newacronym{acf}{ACF}{autocorrelation function}
\newacronym{dsc}{DSC}{differential scanning calorimetry}
\newacronym{fsc}{FSC}{fast scanning calorimetry}
\newacronym{mcmc}{MCMC}{Markov-chain Monte Carlo}
\newacronym{md}{MD}{molecular dynamics}
\newacronym{msd}{MSD}{mean squared displacement}
\newacronym{vft}{VFT}{Vogel-Fulcher-Tammann}

\DeclareSIUnit\angstrom{\text{Å}}
\DeclareSIUnit\bar{bar}

\newcommand{\addchalmersphysics}{Department of Physics, Chalmers University of Technology, SE-41296, Gothenburg, Sweden}
\newcommand{\addchalmerschemistry}{Department of Chemistry and Chemical Engineering, Chalmers University of Technology, SE-41296, Gothenburg, Sweden}

\newcommand{\papertitle}{Probing Glass Formation in Perylene Derivatives via Atomic Scale Simulations and Bayesian Regression}

\begin{document}

\title{\papertitle{}}
\hypersetup{pdfauthor={Eric Lindgren, Jan Swensson, Christian Müller, and Paul Erhart}}
\hypersetup{pdftitle={\papertitle{}}}

\author{Eric Lindgren}
\author{Jan Swenson}
\affiliation{\addchalmersphysics}
\author{Christian Müller}
\affiliation{\addchalmerschemistry}
\author{Paul Erhart}
\email{erhart@chalmers.se}
\affiliation{\addchalmersphysics}

\begin{abstract}
While the structural dynamics of chromophores are of interest for a range of applications, it is experimentally very challenging to resolve the underlying microscopic mechanisms.
Glassy dynamics are also challenging for atomistic simulations due to the underlying dramatic slowdown over many orders of magnitude.
Here, we address this issue by combining atomic scale simulations with autocorrelation function analysis and Bayesian regression, and apply this approach to a set of perylene derivatives as prototypical chromophores.
The predicted glass transition temperatures and kinetic fragilities are in semi-quantitative agreement with experimental data.
By analyzing the underlying dynamics via the normal vector autocorrelation function, we are able to connect the $\beta$ and $\alpha$-relaxation processes in these materials to caged (or librational) dynamics and cooperative rotations of the molecules, respectively.
The workflow presented in this work serves as a stepping stone toward understanding glassy dynamics in many-component mixtures of perylene derivatives and is readily extendable to other systems of chromophores.
\end{abstract}

\maketitle

\vspace{1in}


Chromophores are an important class of materials with a range of potential and realized applications in the area of energy conversion thanks to their exceptional optical properties.
Chromophores have been studied, e.g., as active materials in solar cells \cite{ZhaLinQi22, ZhuZhaZho24, HouIngFri18, TanLiaChe24, SnyDeL18}, organic light-emitting diodes \cite{HaHurPat21, ChiPuKid18}, and photoswitchable and solar thermal storage systems \cite{SalMot24, WanHölMot22, UjiZähKer23}.
The properties of these materials are sensitive to both the structural arrangements of the molecules and their dynamic behavior.
The dynamics as manifested in macroscopic properties such as viscosity and diffusivity, are also important for solution processing, which is currently the most common approach for large-scale manufacturing of devices based on these materials.
Controlling viscosity and diffusivity is often achieved through glass formation \cite{Mül15}, which can occur upon rapid cooling, bypassing crystallization and resulting in a glassy state that lacks long-range order.
The glass transition is characterized by a dramatic slow down in the materials dynamics over a narrow temperature range that is commonly probed via the temperature dependence of, e.g., the viscosity (via rheometry), the density (via dilatometry) or the heat exchanged with the environment (via calorimetry).

\begin{figure}
    \centering
    \includegraphics[width=\linewidth]{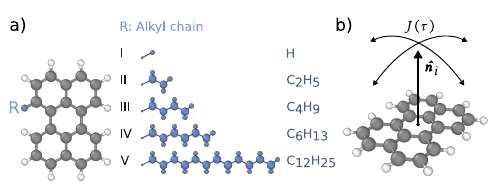}
    \caption{
        (a) Perylene derivatives studied in this work.
        (b) Schematic representation of the normal vector \acrlong{acf} $J(\tau)$, see \autoref{eq:autocorrelation}.
    }
    \label{fig:molecules-and-models}
\end{figure}

For practical use, it is crucial to achieve glass formation controllably at modest cooling rates.
In this context, using mixtures of perylene derivatives, it has been shown that increasingly stronger glass formers can be systematically obtained by increasing the number of components.
This principle works even though the underlying molecules are weak (``fragile'') glass formers in single-component systems \cite{HulCraKus21}.
Moreover, it has been found that such many-component mixtures have further benefits, including significantly improved thermal stability \cite{PalHulHan23}.
While many-components mixtures thus have very high potential for materials design, the much larger design space also renders understanding the underlying dynamical processes much more challenging.
Here, as a first step toward a systematic understanding of these materials, we investigate glass formation in \emph{single}-component liquids of perylene derivatives (\autoref{fig:molecules-and-models}a) using \gls{md} simulations in combination with Bayesian regression.

While the microscopic dynamics of glass-forming systems can be explored via \gls{md} simulations \cite{Wah91, LewWah93}, it is usually impossible to directly access the temperature range in which the glass transition occurs due the time-scale limitations of this technique.
Here, to extend the temperature range, we combine \gls{md} simulations with Bayesian regression, which allows us to predict glass transition temperatures as well as the propensity for glass formation (expressed via the kinetic fragility).
To this end, we observe the temperature dependence of the dynamics via the diffusivity, which is anticorrelated with the viscosity but computationally easier to converge than the latter.
Our results for the glass transition temperatures and the kinetic fragility are in semi-quantitative agreement with experimental data, supporting the viability of the simulation approach.
To gain insight into the microscopic processes we analyze the time-\gls{acf} of the molecular orientation, which reveals three distinct dynamic regimes corresponding to intramolecular motion as well as $\beta$ and $\alpha$-relaxation processes.
Our work thereby establishes the viability of this simulation approach and lays the groundwork for future studies of the evolution of the dynamics in many-component mixtures.


We considered five perylene derivatives (\autoref{fig:molecules-and-models}a), which differ with respect to the length $n$ of the pendant alkyl chain  \ce{C_nH_{2n+1}} attached to one of the bay positions.
Monomer \textbf{I} corresponds to regular perylene with no alkyl chain, whereas monomers \textbf{II}--\textbf{V} have alkyl chains containing two ($n=2$), four ($n=4$), six ($n=6$), and twelve ($n=12$) carbon atoms, respectively. 


\begin{figure*}
    \centering
    \includegraphics{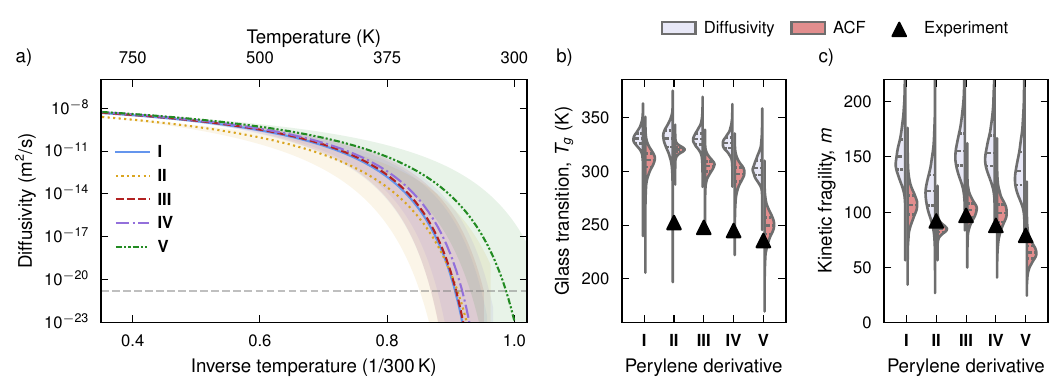}
    \caption{
    (a) Extrapolation of the temperature dependence of the \gls{vft} fit of the diffusivity to lower temperatures (see \autoref{si:d-from-msd} in the Supplementary Information for the \gls{md} data).
    The glass transition temperature $T_g$ is defined as the temperature where the \gls{msd} over \qty{100}{\second} reaches \qty{100}{\angstrom}, denoted by the horizontal gray dashed line. 
    The error band corresponds to one standard deviation.
    (b,c) Violin plots of (b) the glass transition temperature $T_g$ and (c) the kinetic fragility $m$ estimated from both the diffusivity and the normal vector \gls{acf}.
    Experimental values are from Ref.~\citenum{HulCraKus21}.
    $T_g$ values were experimentally obtained from \gls{dsc} first heating thermograms with a heating rate of \qty{0.17}{\kelvin \cdot \second^{-1}}.
    The kinetic fragility was obtained from \gls{fsc} measurements for various cooling rates as $m = - d \log{|q|} / d(T_g/T_f' ) |_{T_f' = T_g}$, where $q$ is the cooling rate and $T_f'$ is the fictive \gls{fsc} temperature.
    The simulated values for the kinetic fragility were computed from the \gls{vft} parameters as $m = B \, T_g / \left[ \ln(10) \cdot (T_g - T_\text{VF})^2)\right]$.
    Note that $T_g$ is typically not observed experimentally for derivative \textbf{I}, due to its strong tendency to crystallize.
    }
    \label{fig:glass_transition}
\end{figure*}

To characterize the dynamics of the perylene derivatives, we first consider the molecular diffusivity $D$, which can be obtained from the \gls{msd} $\left<\Delta r^2\right>$ of the molecular centroid positions \cite{FreSmi01},
\begin{equation}
    \left<\Delta r^2\right> = 6 D \tau
    \label{eq:diffusivity}
\end{equation}
The diffusivity was computed using production runs with a duration of up to \qty{10}{\nano\second}.

To obtain more detailed insight into the underlying microscopic properties, we also analyzed the \gls{acf} of the normal vectors indicating the orientation of each individual molecule (\autoref{fig:molecules-and-models}b) given by
\begin{equation}
   J(\tau) = \left\langle \boldsymbol{\hat{n}}_i(t) \cdot \boldsymbol{\hat{n}}_i(t+\tau) \right\rangle_{it}
   \label{eq:autocorrelation}
\end{equation}
Here, $\boldsymbol{\hat{n}}_i(t)$ is the normal vector of molecule $i$ (\autoref{si:compute-nacf}).
The ensemble average applies over all times $t$ and each molecule $i$ in the system.
Equation \eqref{eq:autocorrelation} can be efficiently evaluated using the Wiener-Kinchin theorem.

One may also extract the standard error as an uncertainty estimate for $J(\tau)$ from the \gls{acf} for each molecule $J_i(\tau)$ before computing the ensemble average in \autoref{eq:autocorrelation} according to
\begin{equation}
    \sigma_{J}(\tau) = \sqrt{\textrm{Var}\left( \left\{ J_i \right\}_{i=1}^N \right) \bigg/ N},
    \label{eq:acf-var}
\end{equation}
where $N$ is the number of molecules in the system.

Since the \glspl{acf} spans multiple orders of magnitude in time, production runs of different length were conducted.
To sample short and long-time scales, simulations with a length of respectively \qty{100}{\pico \second} and \qty{10}{\nano \second} were carried out with snapshots being written every \qty{1}{\femto \second} and \qty{100}{\femto\second}, respectively (\autoref{si:compute-nacf}).
The normal vector \glspl{acf} were calculated for both production runs and subsequently spliced together at a time lag of \qty{1}{\pico \second}.



We begin by analyzing the temperature dependence of the molecular diffusivity (\autoref{fig:glass_transition}a; also see \autoref{si:d-from-msd}).
When obtaining these data from \gls{md} simulations we are limited by the time scale that is reachable via the latter.
While one can reach on the order of \qty{1}{\micro\second} in total simulation length, one has to keep in mind that computing the diffusivity via the \gls{msd} according to \autoref{eq:diffusivity} requires oversampling.
For the present systems, this implies that the \gls{msd} can no longer be reliably obtained for temperatures below approximately \qty{400}{\kelvin}.
This is below the experimental melting point of pure perylene of around \qty{550}{\kelvin} but above the experimental glass transition temperatures, which range around \qty{250}{\kelvin} \cite{HulCraKus21}.

In order to be able to gain information about the behavior at these temperatures, we need to extrapolate.
However, since the diffusivity and other properties change rapidly over many orders of magnitude in this region, this extrapolation must be done with care and account for error propagation.
To this end, we employ the \gls{vft} equation and combine it with Bayesian regression.
The former describes the temperature dependence of, e.g., the viscosity or the diffusivity of fragile glass formers, allowing for non-Arrhenius behavior.
While the \gls{vft} equation is empirical in nature, it is widely used in the analysis of glass-forming systems and provides an accurate fit for many experimental observations as well as the data obtained here (\autoref{fig:glass_transition}a).
For the diffusivity it reads
\begin{align}
    D(T) = D_0 \exp\left[-B/\left(T - T_\text{VF}\right)\right],
    \label{eq:vft-diffusivity}
\end{align}
where $D_0$ is a prefactor, $T_\text{VF}$ is the Vogel-Fulcher temperature, and $B$ is a parameter akin to a pseudo-activation energy \cite{Rau00}.
The parameters of the \gls{vft} equation can in turn be used to compute the kinetic fragility $m = B \, T_g / \left[ \ln(10) \cdot (T_g - T_\text{VF})^2)\right]$, where $T_g$ is the estimated glass transition temperature \cite{PipAbdXu23, NgaFloPla02}.

Due to the exponential in \autoref{eq:vft-diffusivity} extrapolation and error propagation require care, which we handle here via Bayesian regression.
The latter is a technique in which a model $M(\boldsymbol{\theta})$ with parameters represented by a parameter vector $\boldsymbol{\theta}=[D_0, T_\text{VF}, B]$ is fitted to a set of data $\mathcal{D}$ given prior information $\mathcal{I}$, using Bayes' theorem,
\begin{align}
    p(\boldsymbol{\theta} | \mathcal{D}, \mathcal{I}) \propto p(\mathcal{D} | \boldsymbol{\theta}, \mathcal{I}) p(\boldsymbol{\theta} | \mathcal{I}) 
    \label{eq:bayes}
\end{align}
The advantage of a Bayesian approach is twofold.
First, prior beliefs are clearly stated in the prior distribution $p(\boldsymbol{\theta} | \mathcal{I})$.
Second, error estimates are readily extractable from the posterior distribution $p(\boldsymbol{\theta} | \mathcal{D}, \mathcal{I})$, since data uncertainties and errors can be encoded in the likelihood function $p(\mathcal{D} | \boldsymbol{\theta}, \mathcal{I})$.
We then sample the posterior distribution $p(\boldsymbol{\theta} | \mathcal{D}, \mathcal{I})$ via \gls{mcmc} simulations using the diffusivity data from \gls{md} simulations to fit the \gls{vft} equation  (see \autoref{si:bayesian-fitting} for details).
This allows us to extrapolate the diffusivity to lower temperatures along with controlled error estimates (\autoref{fig:glass_transition}a).

The temperature at which the system transitions into a glassy state is denoted by the glass transition temperature $T_g$.
$T_g$ cannot be uniquely defined but is rather set by a pragmatic property-dependent threshold.
For example, one often takes $T_g$ as the temperature where the viscosity reaches a value of \qty{e11}{\pascal\cdot\second} \cite{NgaFloPla02}.
In the present work, when considering the diffusivity, we adopt a threshold of \qty{17e-22}{\meter^2\per\second}, which corresponds to a \gls{msd} of \qty{100}{\angstrom^2} over \qty{100}{\second}.
In other words, it specifies the onset of diffusion beyond the first-nearest neighbor shell.
We emphasize that since the viscosity and similarly the diffusivity change very steeply around the glass transition (\autoref{fig:glass_transition}a) the threshold value has only a modest effect on the values obtained for $T_g$.
For example, increasing or decreasing the threshold by two orders of magnitude changes our estimates for $T_g$ by only \qty{+-5}{\kelvin}

The glass transition temperatures obtained here are in semi-quantitative agreement with experiments, and correctly predict the trend from \textbf{II} to \textbf{V} \cite{HulCraKus21} (\autoref{fig:glass_transition}b).
However, the simulated $T_g$ values are overestimated compared to experimental values obtained by \acrfull{dsc} by \qtyrange{50}{70}{\kelvin}.
The predicted kinetic fragilities are also in agreement with experimentally obtained values from \acrfull{fsc} (\autoref{fig:glass_transition}c).

The glass transition temperature decreases systematically with increasing alkyl chain length.
Conceptually, this can be explained by an increase in the effective volume available to each molecule caged by its neighbors, due to the longer pendant groups.
It is, however, noteworthy that the kinetic fragility exhibits a maximum for \textbf{III}, which features a butyl pendant chain --- a non-trivial behavior that is observed in both experiment and simulation.

\begin{figure*}
    \centering
    \includegraphics{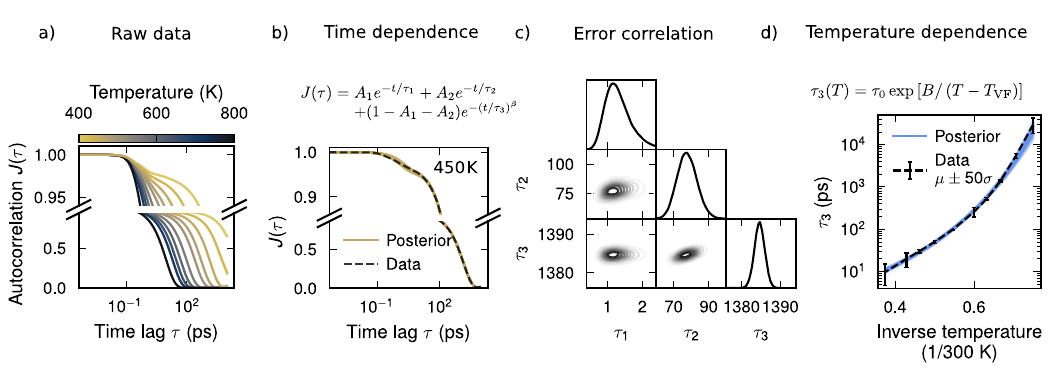}
    \caption{
    Bayesian regression workflow used to extrapolate the normal vector \gls{acf} to longer timescales.
    (a) Normal vector \glspl{acf} for perylene derivative \textbf{I} at different temperatures.
    Note that the y-axis has been split using different scales to reveal the multiple steps in the \gls{acf}.
    (b) Normal vector \gls{acf} for perylene derivative \textbf{I} at \qty{45}{\kelvin} along with the corresponding posterior distribution of fits to \autoref{eq:triple-exponential}.
    (c) Subset of the posterior distribution in (b), $p(\tau_1, \tau_2, \tau_3 | \mathcal{D}, \mathcal{I})$, shown as a corner plot.
    (d) Fit to the \gls{vft} expression \autoref{eq:vft-acf} using the mean $\mu$, and standard deviation $\sigma$ of the marginal distribution $p(\tau_3 | \mathcal{D}, \mathcal{I})$.
    }
    \label{fig:workflow}
\end{figure*}

We now turn to the normal vector \gls{acf} $J(\tau)$ (\autoref{eq:autocorrelation}) to gain further insight into the relaxation processes close to the glass transition (\autoref{fig:workflow}a).
We demonstrate the procedure for obtaining the temperature dependence of $J(\tau)$ for derivative \textbf{I}, noting that the general temperature dependence of $J(\tau)$ is consistent for all perylene derivatives \textbf{I}--\textbf{V} (see Supplementary Information \autoref{si:all-acfs} for the \glspl{acf} for all perylene derivatives).

First, we observe that the correlation time of $J(\tau)$ depends strongly on temperature, ranging from \qty{100}{\pico\second} at \qty{800}{\kelvin} to $>\qty{10}{\nano\second}$ at \qty{400}{\kelvin}.
At \qty{800}{\kelvin} the perylene molecules thus maintain their orientation over a time scale on the order of \qty{100}{\pico\second}, while they are effectively locked in their orientation over \qty{10}{\nano\second} at \qty{400}{\kelvin}.

Second, the \glspl{acf} can be described by the sum of two exponential functions and one stretched exponential function, where the latter is a common feature of correlation functions in glassy systems \cite{LewWah94}
\begin{equation}
    J(\tau) = A_1 e^{-\tau/\tau_1} + A_2 e^{-\tau/\tau_2} + (1-A_1 - A_2) e^{(-\tau/\tau_3)^\beta}
    \label{eq:triple-exponential}
\end{equation}
The timescales $\tau_1$, $\tau_2$, and $\tau_3$ are separated by several orders of magnitude at low temperatures, with $\tau_1 \approx \qty{0.1}{\pico\second}$, $\tau_2 \approx \qty{10}{\pico\second}$, and $\tau_3 \approx \qty{1}{\nano\second}$ at \qty{450}{\kelvin}.
$\beta \leq 1$ is the stretch exponent for the long timescale component.

\begin{figure}
    \centering
    \includegraphics{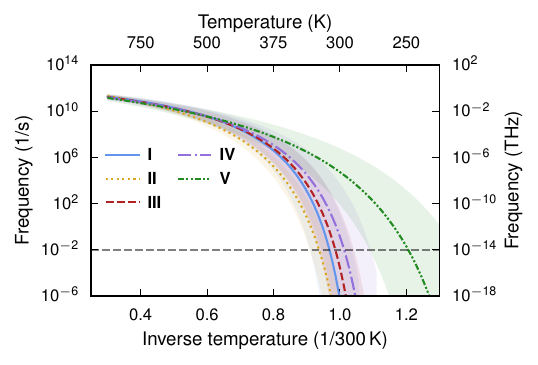}
    \caption{
    Extrapolation of the temperature dependence of the slowest process represented by $\tau_3(T)$, to lower temperatures and thus lower frequencies.
    The error band corresponds to $\pm$ one standard deviation.
    $T_g$ is defined to be the temperature which the timescale reaches \qty{100}{\second}, represented by the horizontal gray dashed line.
    }
    \label{fig:acf-extrapolation}
\end{figure}

We can apply the same Bayesian regression workflow as for the diffusivity to estimate the glass transition temperature and kinetic fragility from the temperature dependence of the normal vector \gls{acf}.
However, an additional step is required compared to the diffusivity, as the normal vector \gls{acf} needs to be fitted to \autoref{eq:triple-exponential} for each temperature (\autoref{fig:workflow}b).
Each fit yields a full posterior probability distribution $p(A_1, A_2, \tau_1, \tau_2, \tau_3, \beta| T, \mathcal{D}, \mathcal{I})$.
An estimate for the timescale of the slowest process captured by the \gls{acf} presented by $\tau_3$ with uncertainty estimates can then be obtained from the marginal distribution $p(\tau_3 | T, \mathcal{D}, \mathcal{I})$ for each temperature (\autoref{fig:workflow}c). 
A \gls{vft} equation of the form
\begin{align}
    \tau_3(T) = \tau_0 \exp\left[B/\left(T - T_\text{VF}\right)\right],
    \label{eq:vft-acf}
\end{align}
is then fitted to the temperature dependence of $\tau_3$, which allows for a similar extrapolation to longer timescales as in the case of the diffusivity (\autoref{fig:acf-extrapolation}).
Here, the threshold for $\tau_3$ above which the system is deemed to be in a glassy state was again taken to be \qty{100}{\second} \cite{NgaFloPla02}.
Note that the resulting glass transition temperature is relatively insensitive to this particular threshold, as increasing or decreasing the threshold by two orders of magnitude only changes $T_g$ by \qty{+-7}{\kelvin}.

The estimates for both the glass transition temperature and the kinetic fragility from the normal-vector \gls{acf} and the diffusivity generally agree with each other (\autoref{fig:glass_transition}b,c).
Both schemes capture the trend of decreasing $T_g$ with increasing length of the alkyl chain of the perylene derivative. 
However, the estimates from the diffusivity are higher than those from the normal-vector \gls{acf} typically by \qtyrange{10}{30}{\kelvin} for the glass transition temperature and by \numrange{10}{40} for the kinetic fragility.
This difference is due to the two observables probing different processes.
The diffusivity is sensitive to the diffusion of the monomers, while the normal vector \gls{acf} probes the rotational motion of the monomer.
The normal vector \gls{acf} and the diffusivity are thus complementary.
The difference in $T_g$ between both observables suggests that the monomers continue to rotate on long timescales \qtyrange{10}{30}{\kelvin} below the temperature at which diffusion has slowed down. 

\begin{figure}
    \centering
    \includegraphics{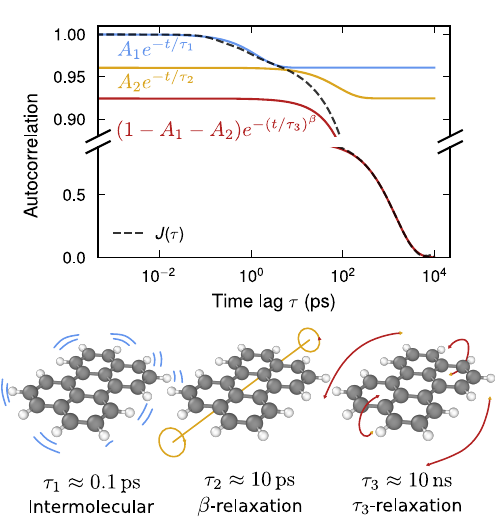}
    \caption{
    Decomposition of the normal vector \gls{acf} $J(\tau)$ into individual exponential functions representing three different relaxation processes.
    The fastest process with a correlation time of about \qty{0.1}{\pico\second} at \qty{450}{\kelvin} corresponds to intramolecular atomic motion.
    The second one ($\beta$-relaxation) with a correlation time of approximately \qty{10}{\pico\second} at \qty{450}{\kelvin} corresponds to librational motion and twisting of the perylene core.
    Finally, the slowest process ($\tau_3$-relaxation) with a correlation time of approximately \qty{10}{\nano\second} at \qty{450}{\kelvin} corresponds to the hindered rotational reorientation of the perylene molecules due to intermolecular interactions.
    }
    \label{fig:timescales-schematic}
\end{figure}

We can elucidate the relaxation processes in the system by decomposing the \gls{acf} into the contribution of each exponential function that make up $J(\tau)$ (\autoref{fig:timescales-schematic}).
The separation of timescales between the processes allows the selective application of frequency filters in the Fourier domain, corresponding to the timescales represented by $\tau_1$, $\tau_2$, and $\tau_3$.
These filters are applied to the trajectory of a single perylene molecule extracted from the entire \gls{md} trajectory, and allows us to single out the dynamics that correspond to each process (see the supplementary movie for a visual representation of this scheme, and \autoref{si:decomposed-autocorrelation} of the Supplementary Information for further details).

We study the dynamics of perylene derivative \textbf{I} at a temperature of \qty{450}{\kelvin} as an example of this scheme (\autoref{fig:timescales-schematic}).
The fastest process with time scale $\tau_1$ corresponds to intramolecular atomic motion.
The second fastest process, $\tau_2$, corresponds to $\beta$-relaxation enforced by caging by neighboring molecules, such as libration and twisting of the perylene core.
Neither the $\tau_1$ nor the $\tau_2$ processes significantly affect the orientation of the molecule, as is evident by their small amplitude.
The bulk of the autocorrelation function $J(\tau)$ is made up of the slow $\tau_3$ process.
$\tau_3$ corresponds to cooperative intermolecular processes, such as reorientation of molecules.
The reorientation of a molecule requires neighboring molecules to rotate, which takes place over rapidly increasing timescales as the temperature is decreased.
The experimental and simulated values of the kinetic fragility indicate that all derivatives studied in this work are fragile glass formers.
In fragile glass formers, the level of cooperation decreases significantly at temperatures greater than $T_g$ \cite{SaiSaiGre06}.
That $\tau_3$ captures cooperative reorientation even in the supercooled regime at \qty{450}{\kelvin} thus indicates that it is sensitive to processes that are more prominent close to the glass transition in fragile glass formers.
Based on this, we attribute $\tau_3$ to be related to $\alpha$-relaxation, and that the microscopic mechanism driving glass formation in perylene derivatives \textbf{I}--\textbf{V} is the cooperative reorientation of the molecules.

Given the sources of uncertainty related to the underlying empirical force field used in the \gls{md} simulations and the extrapolation over many orders of magnitude, we consider the overall agreement of the predicted glass transition temperatures and kinetic fragilities with the experimental data very encouraging.
The normal vector \gls{acf} in particular show semi-quantiative agreement with experiments, with the \gls{acf} systematically yielding both lower glass transition temperatures and kinetic fragilities than the diffusivity (\autoref{fig:glass_transition}b,c).
This difference highlights the complementarity of the diffusivity and the normal vector \gls{acf}, as they are more sensitive to molecular diffusion and rotation, respectively.
The estimated higher value of the glass transition temperature from the diffusivity can be understood as molecular diffusion freezing in at a higher temperature compared to rotation.
The processes driving glass formation are thus cooperative rotational processes, as elucidated by the decomposition of the normal vector \gls{acf}.
This is supported by the large kinetic fragility deduced for all derivatives (\autoref{fig:timescales-schematic}).
Capturing both diffusion and rotation is hence key in order to accurately describe the relaxation processes in the fragile perylene derivatives studied in this work.

Both the normal vector \gls{acf} and the diffusivity systematically overestimate the glass transition temperature and the kinetic fragility compared to experiment.
The overestimation of the kinetic fragility suggests that the processes represented by $\tau_3$ in the \gls{md} simulation are slower than those encountered during experiments.
This could be caused by the intermolecular interactions in the simulation being somewhat too soft, which would point toward a limitation in the accuracy of the underlying force field.
Another possible explanation could be that the normal vector \gls{acf} overestimates the time scale of processes in the system.

As noted in the introduction, experimentally the glass transition can also be detected as a change in the thermal expansion of the material, an approach that is also occasionally adopted in simulations \cite{PatDieBro16, ReiJavMan21, CalBomMas22, FarEbr24, MarUraDu24}.
It is therefore instructive to contrast this approach with the one based on diffusivity and time \glspl{acf} used in the present work.
For the present systems we observe a change in the thermal expansion coefficient at a temperature of around \qty{400}{\kelvin}, which would suggest much higher glass transition temperatures (\autoref{si:tg-from-annealing}).
At the same time, one can observe from the analysis of the normal vector \gls{acf} that in this temperature range the relaxation time for the slowest process $\tau_3$ reaches the limit of the \gls{md} time scale.
The change in the thermal expansion is thus merely a direct result of the latter rather than a feature of the system.
As a result, at least for the present systems the analysis of the thermal expansion cannot be expected to yield a physically meaningful estimate of the glass transition temperature.

The method for extending the temperature range using Bayesian regression presented in this work allows us to study relaxation processes in liquid and supercooled liquid systems containing hundreds to thousands of molecules.
Hence, it is possible to make material-specific predictions for the glass transition temperature and the kinetic fragility.
The general approach is directly extendable to other systems, where especially the diffusivity can be readily computed.
This work serves as a first step towards accurately describing the complex relaxation processes in multi-component mixtures of perylene derivatives.
Insight into these relaxation processes is key in obtaining a systematic understanding of the dynamics of perylene derivatives, enabling the design of stronger and more stable glass forming system.

\section*{Computational details}

\Gls{md} simulations were performed using the \textsc{gromacs} package \cite{AbrMurSch15} with the OPLS all-atom force field \cite{JorTir88}.
Topology and structure files where generated using the LigParGen server \cite{JorTir05, DodVilTir17, DodCabTir17}, starting from structures from the automated topology builder and repository \cite{MalZuoBre11, CanEl-Poo13, KozStrMal14}.
A time step of \qty{1}{\femto\second} was used for all simulations, in combination with constraining the hydrogen atoms using the linear constraint solver algorithm \cite{HesBekBer97}.
The simulation cell contained between 500 and 2000 molecules depending on the length of the alkyl chain of the perylene derivative, and simulations were performed at temperatures in the range \qtyrange{400}{800}{\kelvin}.

Each system was equilibrated at the target temperature prior to production using the following protocol.
First, the system energy was minimized using a steepest descent optimizer, after which a simulation of \qty{1}{\nano\second} was performed in the NVT ensemble.
This was followed by a \qty{1}{\nano\second} run in the NPT ensemble at a pressure of \qty{2}{\kilo\bar} using a Berendsen barostat \cite{BerPosvan84} to avoid cavitation.
The high-pressure NPT simulation was followed by a \qty{10}{\nano\second} NPT simulation at \qty{1}{\bar}.
Finally, production runs were carried out in the NPT ensemble using the stochastic pressure-rescaling barostat and a stochastic velocity-rescaling thermostat \cite{BusDonPar07} to obtain the diffusivity as well as the short and long-time normal vector \gls{acf} (\autoref{si:compute-nacf}).
The production runs were \qty{100}{\pico \second} and \qty{10}{\nano \second} long, and trajectory files were written every \qty{1}{\femto \second} and \qty{100}{\femto \second}, respectively.

The trajectories resulting from the simulations were then parsed using the \textsc{mdtraj} package \cite{McGBeaHar15} and analyzed using \textsc{python} scripts to compute the correlation function defined by \autoref{eq:autocorrelation}.
Bayesian regression analysis was performed using the \textsc{numpy} \cite{HarMilvan20}, \textsc{pandas} \cite{Mck10, The23}, \textsc{scipy} \cite{VirGomOli20} and \textsc{emcee} \cite{ForHogLan13} packages.
Plots were generated using \textsc{matplotlib} \cite{Hun07}, \textsc{seaborn} \cite{Was21}, \textsc{corner} \cite{For16}, and color maps from \textsc{perfect-cmaps} \cite{Ulm24}.

\section*{Acknowledgments}

We are grateful to Göran Wahnström for helpful discussions.
This work was funded by the Swedish Foundation for Strategic Research via the SwedNESS graduate school (GSn15-0008), the Swedish Research Council (grant numbers 2019-04020, 2020-04935, 2021-05072, 2022-02977), and the Chalmers Initiative for Advancement of Neutron and Synchrotron Techniques.
The computations were enabled by resources provided by the National Academic Infrastructure for Supercomputing in Sweden (NAISS) at C3SE, NSC, and PDC partially funded by the Swedish Research Council through grant agreements no. 2022-06725 and no. 2018-05973.

\end{document}